\pdfoutput=1
\documentclass[12pt,a4paper]{article}
\usepackage{comment}
\usepackage{ifthen}
\usepackage{booktabs}
\newboolean{pdflatex}
\setboolean{pdflatex}{true}

\newboolean{articletitles}
\setboolean{articletitles}{true}

\newboolean{uprightparticles}
\setboolean{uprightparticles}{false}

\def\paperauthors{LHCb collaboration}
\def\paperasciititle{Identification of charm jets at LHCb}
\def\papertitle{Identification of charm jets \\ at \lhcb}
\def\paperkeywords{{High Energy Physics}, {LHCb}, {Jet Tagging}, {Performance}}
\def\papercopyright{\the\year\ CERN for the benefit of the LHCb collaboration}
\def\paperlicence{CC BY 4.0 licence}
\def\paperlicenceurl{https://creativecommons.org/licenses/by/4.0/}

\def\ptj{\ensuremath{\pt(j)}\xspace}

\usepackage[top=1in, bottom=1.25in, left=1in, right=1in]{geometry}

\usepackage{microtype}
\usepackage{lineno}
\usepackage{xspace}
\usepackage{caption}

\usepackage{graphicx}
\usepackage{color}
\usepackage{colortbl}

\usepackage{amsmath}
\usepackage{amssymb}
\usepackage{amsfonts}
\usepackage{upgreek}

\newcommand*\patchAmsMathEnvironmentForLineno[1]{
\expandafter\let\csname old#1\expandafter\endcsname\csname #1\endcsname
\expandafter\let\csname oldend#1\expandafter\endcsname\csname
end#1\endcsname
 \renewenvironment{#1}
   {\linenomath\csname old#1\endcsname}
   {\csname oldend#1\endcsname\endlinenomath}
}
\newcommand*\patchBothAmsMathEnvironmentsForLineno[1]{
  \patchAmsMathEnvironmentForLineno{#1}
  \patchAmsMathEnvironmentForLineno{#1*}
}
\AtBeginDocument{
\patchBothAmsMathEnvironmentsForLineno{equation}
\patchBothAmsMathEnvironmentsForLineno{align}
\patchBothAmsMathEnvironmentsForLineno{flalign}
\patchBothAmsMathEnvironmentsForLineno{alignat}
\patchBothAmsMathEnvironmentsForLineno{gather}
\patchBothAmsMathEnvironmentsForLineno{multline}
\patchBothAmsMathEnvironmentsForLineno{eqnarray}
}

\usepackage{hyperxmp}

\usepackage[pdftex,
            pdfauthor={\paperauthors},
            pdftitle={\paperasciititle},
            pdfkeywords={\paperkeywords},
            pdfcopyright={Copyright (C) \papercopyright},
            pdflicenseurl={\paperlicenceurl}]{hyperref}

\usepackage[colorinlistoftodos,textsize=scriptsize]{todonotes}

\usepackage[bottom,flushmargin,hang,multiple]{footmisc}

\usepackage[all]{hypcap}

\usepackage{xspace}
\usepackage{upgreek}

\def\lhcb   {\mbox{LHCb}\xspace}

\def\MagUp {\mbox{\em Mag\kern -0.05em Up}\xspace}

\ifthenelse{\boolean{uprightparticles}}
{

 \def\Ppi         {\ensuremath{\uppi}\xspace}

 \def\Ppsi        {\ensuremath{\uppsi}\xspace}

 \def\PDelta      {\ensuremath{\Delta}\xspace}
 \def\PXi         {\ensuremath{\Xi}\xspace}
 \def\PLambda     {\ensuremath{\Lambda}\xspace}
 \def\PSigma      {\ensuremath{\Sigma}\xspace}
 \def\POmega      {\ensuremath{\Omega}\xspace}
 \def\PUpsilon    {\ensuremath{\Upsilon}\xspace}

 \def\PB      {\ensuremath{\mathrm{B}}\xspace}
 
 \def\PD      {\ensuremath{\mathrm{D}}\xspace}

 \def\PJ      {\ensuremath{\mathrm{J}}\xspace}
 \def\PK      {\ensuremath{\mathrm{K}}\xspace}

 \def\PW      {\ensuremath{\mathrm{W}}\xspace}

 \def\PZ      {\ensuremath{\mathrm{Z}}\xspace}
 
 \def\Pb      {\ensuremath{\mathrm{b}}\xspace}
 \def\Pc      {\ensuremath{\mathrm{c}}\xspace}

 \def\Pi      {\ensuremath{\mathrm{i}}\xspace}

 \def\Pp      {\ensuremath{\mathrm{p}}\xspace}

 \def\Ps      {\ensuremath{\mathrm{s}}\xspace}

 \def\thebaroffset{0.0em}
}
{

 \def\Ppi         {\ensuremath{\pi}\xspace}

 \def\Ppsi        {\ensuremath{\psi}\xspace}
 
 \mathchardef\PDelta="7101
 \mathchardef\PXi="7104
 \mathchardef\PLambda="7103
 \mathchardef\PSigma="7106
 \mathchardef\POmega="710A
 \mathchardef\PUpsilon="7107
 
 \def\PB      {\ensuremath{B}\xspace}
 
 \def\PD      {\ensuremath{D}\xspace}

 \def\PJ      {\ensuremath{J}\xspace}
 \def\PK      {\ensuremath{K}\xspace}

 \def\PW      {\ensuremath{W}\xspace}

 \def\PZ      {\ensuremath{Z}\xspace}
 
 \def\Pb      {\ensuremath{b}\xspace}
 \def\Pc      {\ensuremath{c}\xspace}

 \def\Pi      {\ensuremath{i}\xspace}

 \def\Pp      {\ensuremath{p}\xspace}

 \def\Ps      {\ensuremath{s}\xspace}

 \def\thebaroffset{0.18em}
}
\newcommand{\offsetoverline}[2][\thebaroffset]{\kern #1\overline{\kern -#1 #2}}

\makeatletter
\ifcase \@ptsize \relax
  \newcommand{\miniscule}{\@setfontsize\miniscule{4}{5}}
\or
  \newcommand{\miniscule}{\@setfontsize\miniscule{5}{6}}
\or
  \newcommand{\miniscule}{\@setfontsize\miniscule{5}{6}}
\fi
\makeatother

\DeclareRobustCommand{\optbar}[1]{\shortstack{{\miniscule (\rule[.5ex]{1.25em}{.18mm})}
  \\ [-.7ex] $#1$}}

\def\W      {{\ensuremath{\PW}}\xspace}

\def\Z      {{\ensuremath{\PZ}}\xspace}

\def\squark    {{\ensuremath{\Ps}}\xspace}

\def\cquark    {{\ensuremath{\Pc}}\xspace}
\def\cquarkbar {{\ensuremath{\overline \cquark}}\xspace}
\def\ccbar     {{\ensuremath{\cquark\cquarkbar}}\xspace}
\def\bquark    {{\ensuremath{\Pb}}\xspace}
\def\bquarkbar {{\ensuremath{\overline \bquark}}\xspace}
\def\bbbar     {{\ensuremath{\bquark\bquarkbar}}\xspace}

\def\pion   {{\ensuremath{\Ppi}}\xspace}

\def\pip    {{\ensuremath{\pion^+}}\xspace}
\def\pim    {{\ensuremath{\pion^-}}\xspace}

\def\kaon    {{\ensuremath{\PK}}\xspace}

\def\KorKbar {\kern \thebaroffset\optbar{\kern -\thebaroffset \PK}{}\xspace}

\def\Km      {{\ensuremath{\kaon^-}}\xspace}

\def\D       {{\ensuremath{\PD}}\xspace}

\def\DorDbar {\kern \thebaroffset\optbar{\kern -\thebaroffset \PD}\xspace}
\def\Dz      {{\ensuremath{\D^0}}\xspace}

\def\Dp      {{\ensuremath{\D^+}}\xspace}
\def\Dm      {{\ensuremath{\D^-}}\xspace}

\def\DpDm    {\ensuremath{\Dp {\kern -0.16em \Dm}}\xspace}

\def\B       {{\ensuremath{\PB}}\xspace}

\def\BorBbar {\kern \thebaroffset\optbar{\kern -\thebaroffset \PB}\xspace}
\def\Bz      {{\ensuremath{\B^0}}\xspace}

\def\Bd      {{\ensuremath{\B^0}}\xspace}

\def\BdorBdbar {\kern \thebaroffset\optbar{\kern -\thebaroffset \Bd}\xspace}

\def\Bs      {{\ensuremath{\B^0_\squark}}\xspace}

\def\BsorBsbar {\kern \thebaroffset\optbar{\kern -\thebaroffset \Bs}\xspace}

\def\jpsi     {{\ensuremath{{\PJ\mskip -3mu/\mskip -2mu\Ppsi}}}\xspace}

\def\Y#1S{\ensuremath{\PUpsilon{(#1S)}}\xspace}

\def\proton      {{\ensuremath{\Pp}}\xspace}

\def\LorLbar     {\kern \thebaroffset\optbar{\kern -\thebaroffset \PLambda}\xspace}

\newcommand{\decay}[2]{\ensuremath{#1\!\to #2}\xspace}

\def\to                 {\ensuremath{\rightarrow}\xspace}

\def\order   {{\ensuremath{\mathcal{O}}}\xspace}

\def\AT#1     {\ensuremath{A_{\mathrm{T}}^{#1}}\xspace}

\def\C#1      {\ensuremath{\mathcal{C}_{#1}}\xspace}
\def\Cp#1     {\ensuremath{\mathcal{C}_{#1}^{'}}\xspace}
\def\Ceff#1   {\ensuremath{\mathcal{C}_{#1}^{\mathrm{(eff)}}}\xspace}
\def\Cpeff#1  {\ensuremath{\mathcal{C}_{#1}^{'\mathrm{(eff)}}}\xspace}
\def\Ope#1    {\ensuremath{\mathcal{O}_{#1}}\xspace}
\def\Opep#1   {\ensuremath{\mathcal{O}_{#1}^{'}}\xspace}

\newcommand{\nospaceunit}[1]{\ensuremath{\text{#1}}}
\newcommand{\aunit}[1]{\ensuremath{\text{\,#1}}}

\newcommand{\tev}{\aunit{Te\kern -0.1em V}\xspace}
\newcommand{\gev}{\aunit{Ge\kern -0.1em V}\xspace}
\newcommand{\mev}{\aunit{Me\kern -0.1em V}\xspace}
\newcommand{\kev}{\aunit{ke\kern -0.1em V}\xspace}
\newcommand{\ev}{\aunit{e\kern -0.1em V}\xspace}

\newcommand{\mevc}{\ensuremath{\aunit{Me\kern -0.1em V\!/}c}\xspace}
\newcommand{\gevc}{\ensuremath{\aunit{Ge\kern -0.1em V\!/}c}\xspace}
\newcommand{\mevcc}{\ensuremath{\aunit{Me\kern -0.1em V\!/}c^2}\xspace}
\newcommand{\gevcc}{\ensuremath{\aunit{Ge\kern -0.1em V\!/}c^2}\xspace}

\def\mum  {\ensuremath{\,\upmu\nospaceunit{m}}\xspace}

\def\fb   {\ensuremath{\aunit{fb}}\xspace}
\def\invfb   {\ensuremath{\fb^{-1}}\xspace}

\def\order{{\ensuremath{\mathcal{O}}}\xspace}
\newcommand{\chisq}{\ensuremath{\chi^2}\xspace}

\newcommand{\chisqip}{\ensuremath{\chi^2_{\text{IP}}}\xspace}

\def\gsim{{~\raise.15em\hbox{$>$}\kern-.85em
          \lower.35em\hbox{$\sim$}~}\xspace}
\def\lsim{{~\raise.15em\hbox{$<$}\kern-.85em
          \lower.35em\hbox{$\sim$}~}\xspace}

\def\sqs   {\ensuremath{\protect\sqrt{s}}\xspace}

\def\pt         {\ensuremath{p_{\mathrm{T}}}\xspace}

\def\evtgen     {\mbox{\textsc{EvtGen}}\xspace}

\def\geant      {\mbox{\textsc{Geant4}}\xspace}

\def\photos     {\mbox{\textsc{Photos}}\xspace}

\def\pythia     {\mbox{\textsc{Pythia}}\xspace}

\def\tell1  {TELL1\xspace}
\def\ukl1   {UKL1\xspace}

\newcommand{\eg}{\mbox{\itshape e.g.}\xspace}
\newcommand{\ie}{\mbox{\itshape i.e.}\xspace}

\usepackage{cite}
\usepackage{mciteplus}
\usepackage{longtable}

\begin{document}

\renewcommand{\thefootnote}{\fnsymbol{footnote}}
\setcounter{footnote}{1}

\begin{titlepage}
\pagenumbering{roman}

\vspace*{-1.5cm}
\centerline{\large EUROPEAN ORGANIZATION FOR NUCLEAR RESEARCH (CERN)}
\vspace*{1.5cm}
\noindent
\begin{tabular*}{\linewidth}{lc@{\extracolsep{\fill}}r@{\extracolsep{0pt}}}
\vspace*{-1.5cm}\mbox{\!\!\!\includegraphics[width=.14\textwidth]{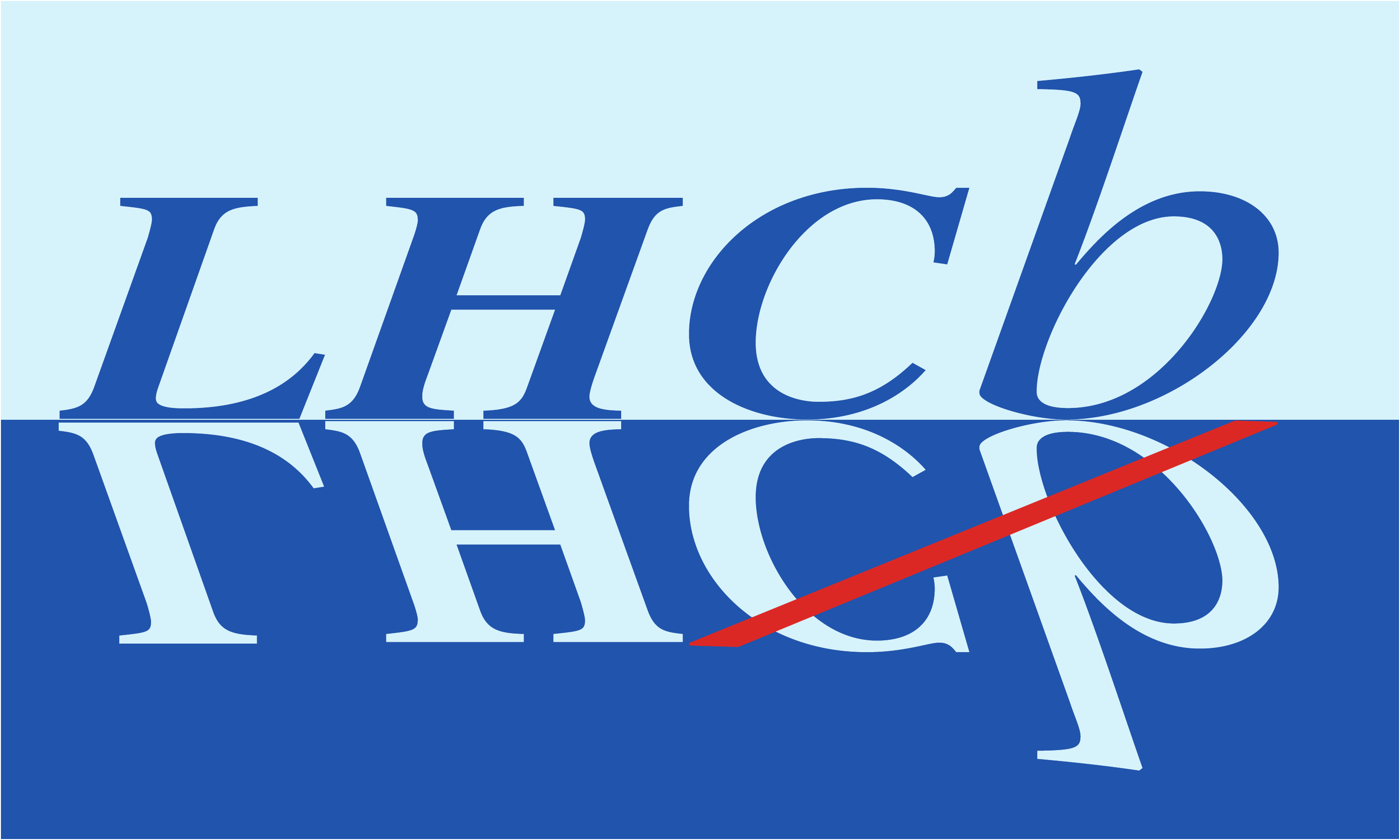}} & &
\\
 & & LHCb-DP-2021-006 \\
 & & 24 February 2022 \\
 & & \\
\end{tabular*}

\vspace*{4.0cm}

{\normalfont\bfseries\boldmath\huge
\begin{center}
  \papertitle
\end{center}
}

\vspace*{2.0cm}

\begin{center}
\paperauthors\footnote{Authors are listed at the end of this paper.}
\end{center}

\vspace{\fill}

\begin{abstract}
  \noindent
  The identification of charm jets is achieved at LHCb for data collected in 2015--2018 using
  a method based on the properties of displaced vertices reconstructed and matched with jets. The performance of this method is determined using a dijet calibration dataset recorded by the LHCb detector and selected such that the jets are unbiased in quantities used in the tagging algorithm. The charm-tagging efficiency is reported as a function of the transverse momentum of the jet. The measured efficiencies are compared to those obtained from simulation and found to be in good agreement.
\end{abstract}

\vspace*{2.0cm}

\begin{center}
  Published in
  the Journal of Instrumentation
\end{center}

\vspace{\fill}

{\footnotesize
\centerline{\copyright~\papercopyright. \href{\paperlicenceurl}{\paperlicence}.}}
\vspace*{2mm}

\end{titlepage}

\newpage
\setcounter{page}{2}
\mbox{~}

\renewcommand{\thefootnote}{\arabic{footnote}}
\setcounter{footnote}{0}

\cleardoublepage

\pagestyle{plain}
\setcounter{page}{1}
\pagenumbering{arabic}

\section{Introduction}
\label{sec:Introduction}

Identification of charm jets, \ie\ those originating from the hadronisation of a charm quark, is of interest in both the study of Standard Model (SM) processes and the search for new physics.
For example,
the production of events containing a \Z boson and a \cquark jet in the forward region provides a direct probe of the charm content of the proton at large parton momentum fractions~\cite{Boettcher:2015sqn,LHCb-PAPER-2021-029}.
Such studies rely upon algorithms capable of distinguishing charm jets from beauty and light-parton jets. An algorithm for identifying both charm and beauty jets reconstructed by the LHCb detector has been used for studies of the dataset recorded in 2011--2012 (Run~1)~\cite{LHCb-PAPER-2015-016}. However, the higher particle multiplicity of 2015--2018 (Run~2) data has been found to degrade its performance~\cite{LHCb-PAPER-2020-018}.
Furthermore, for measurements that only involve charm jets, it is possible to achieve better performance using a dedicated charm-tagging algorithm.

Charm jets are defined as those that have a promptly produced and weakly decaying $c$ hadron with transverse momentum $\pt(c$ hadron$) > 5\gev$ within the jet cone.\footnote{Natural units are used throughout this article.}
Therefore,
the tagging of $c$ jets is performed using displaced vertices (DVs) formed from the decays of such $c$ hadrons.
The choice of using DVs and not single-track or other non-DV-based jet properties, {\em e.g.}\ the number of particles in the jet, is driven by the need for a small misidentification probability of the copious light-parton jets in LHCb $c$-jet analyses.
In addition, the properties of DVs from $c$-hadron decays are known to be well modeled by simulation, which means that only small corrections obtained from control samples are required.
Since DVs can also be formed from the decays of beauty or strange hadrons, or due to artifacts of the reconstruction, the DV-tagged charm yields are obtained by fitting the distributions of DV features with good discrimination power between $c$, $b$, and light-parton jets.

This article presents a dedicated $c$-tagging procedure used to efficiently identify charm jets produced in proton-proton ($pp$) collisions at a centre-of-mass energy  $\sqs=13\tev$ and recorded by the LHCb detector.
The procedure described produces a statistical separation of the different jet-flavour populations resulting in a measurement of the total yield of each category rather than event-by-event flavour tags as given by some other methods. Therefore, the ability to reject light-flavour jets is contained within the uncertainty on the charm-tagging efficiency rather than a mistag rate.
The $c$-tagging efficiency is precisely determined using a sample of unbiased charm-enriched jets obtained from dijet events in a dataset corresponding to an integrated luminosity of $1.7\invfb$ collected in 2016.
This efficiency is reported for jets with $20 < \ptj < 100\gev$ in the pseudorapidity range $2.2 < \eta(j) < 4.2$.
The region below 20\gev is not reported because the $c$-tagging efficiency varies rapidly there, whereas above 100\gev the limited size of the calibration sample prohibits precisely determining the performance.
The $\eta(j)$ range, which was first used in Refs.~\cite{LHCb-PAPER-2015-016,LHCb-PAPER-2015-022,LHCb-PAPER-2015-021}, ensures a nearly uniform $c$-tagging efficiency of about 24\%,
with minimal \ptj or $\eta(j)$ dependence.

\section{Detector and simulation}
\label{sec:Detector}

The \lhcb detector~\cite{LHCb-DP-2008-001,LHCb-DP-2014-002} is a single-arm forward
spectrometer covering
$2<\eta <5$,
designed for the study of particles containing \bquark or \cquark
quarks. The detector includes a high-precision tracking system
consisting of a silicon-strip vertex detector surrounding the $pp$
interaction region, a large-area silicon-strip detector located
upstream of a dipole magnet with a bending power of about
$4{\mathrm{\,Tm}}$, and three stations of silicon-strip detectors and straw
drift tubes placed downstream of the magnet.
The tracking system provides a measurement of the momentum
of charged particles with
a relative uncertainty that varies from 0.5\% at low momentum to 1.0\% at 200\gev.
The minimum distance of a track to a primary $pp$ collision vertex (PV), the impact parameter (IP),
is measured with a resolution of $(15+29/\pt)\mum$,
where \pt is
in\,\gev.
Different types of charged hadrons are distinguished using information
from two ring-imaging Cherenkov detectors.
Photons, electrons, and hadrons are identified by a calorimeter system consisting of
scintillating-pad and preshower detectors, an electromagnetic
and a hadronic calorimeter. Muons are identified by a
system composed of alternating layers of iron and multiwire
proportional chambers.

The online event selection is performed by a trigger,
which consists of a hardware stage, based on information from the calorimeter and muon
systems, followed by a software stage, which applies a full event
reconstruction.
At the hardware trigger stage, events are required to have a muon with high \pt or a hadron, photon, or electron with high transverse energy in the calorimeters. For hadrons, the transverse energy threshold is $3.5\gev$.
The software trigger requires at least one charged particle to be reconstructed with $\pt > 1.6\gev$ that is inconsistent with originating from any PV, as well as the presence of two jets. Both jets are reconstructed as described in Sec.~\ref{sec:identification}, and required to have $\pt>17\gev$. At least one jet is required to have a DV in the jet cone.

Simulation is required to model the effects of the detector acceptance and the imposed selection requirements. In the simulation, $pp$ collisions are generated using \pythia~\cite{Sjostrand:2007gs,*Sjostrand:2006za} with a specific \lhcb configuration~\cite{LHCb-PROC-2010-056}. Decays of unstable particles are described by \evtgen~\cite{Lange:2001uf}, in which final-state radiation is generated using \photos~\cite{davidson2015photos}. The interaction of the generated particles with the detector, and its response, are implemented using the \geant toolkit~\cite{Allison:2006ve, *Agostinelli:2002hh} as described in Ref.~\cite{LHCb-PROC-2011-006}.

\section{Charm-jet identification}
\label{sec:identification}

Jets are reconstructed from particle flow objects~\cite{LHCb-PAPER-2013-058} using the {\sc Fastjet}~\cite{Cacciari:2011ma} implementation of the anti-$k_T$ algorithm~\cite{Cacciari:2008gp} with a jet radius parameter of $R=0.5$. The same jet reconstruction algorithm is used online and offline; however, differences in the reconstruction routines for tracks and calorimeter clusters lead to minor differences between the online and offline jets.
In addition to jets in the fiducial region, offline jets
with ${15 < \ptj < 20\gev}$
are retained for use when unfolding the detector response.

Charm jets are identified based on properties of DVs associated with the jets, reconstructed in a manner similar to that used in Ref.~\cite{LHCb-PAPER-2015-016}. DV candidates are reconstructed using good-quality tracks both within and outside of the jet, with $\pt>0.5\gev$ and $\chisqip>9$, where \chisqip\ is defined as the difference in the vertex-fit \chisq of the PV reconstructed with and without the track under consideration. Tracks are combined into two- and three-body DVs, which are required to form a good-quality vertex, be downstream of the PV, and have an invariant mass greater than $0.4\gev$ and less than that of the $\Bz$ meson.  The corrected mass is required to satisfy
\begin{equation}
m_{\rm cor}({\rm DV}) \equiv \sqrt{m({\rm DV})^2 + [p({\rm DV})\sin{\theta}]^2} +  p({\rm DV})\sin{\theta} > 0.6\gev\,,
\end{equation}
where $\theta$ is the angle between the DV momentum and its direction  of flight, defined by the vector from the $pp$ interaction point to the DV position. In addition, the uncertainty on the corrected mass, as computed from the covariances of the primary and displaced vertices and the DV momentum, is required to be less than $0.5\gev$. Two- and three-body DV candidates that pass these requirements and share one or more tracks are linked together to form $n$-body DVs. All DV candidates are subsequently required to have $\pt>2\gev$ and a significant separation from all PVs.
To reduce backgrounds due to strange-hadron decays and material interactions, DVs are required to have a decay time consistent with a heavy-flavour hadron,
and have a significant separation from all material within and around the vertex detector~\cite{LHCb-DP-2018-002}.

Given that the method presented here is only concerned with tagging charm jets and that DVs with more than four tracks originate predominantly from beauty decays, only DV candidates with two, three, or four tracks are retained.
A DV is associated to a jet when $\Delta R \equiv \sqrt{\Delta \eta^2 + \Delta \phi^2} < 0.5$ between the jet axis and the DV direction  of flight, where $\phi$ denotes the azimuthal angle.
If more than one DV candidate is assigned to a given jet, which occurs for $\mathcal{O}(1\%)$ of charm jets, the candidate with the largest $\pt$ is retained. The key differences between the DV candidates used in this study and those in Ref.~\cite{LHCb-PAPER-2015-016} are a less stringent $\chisqip$ requirement, the addition of the corrected mass uncertainty requirement, and the requirement for no more than four tracks.

To determine the number of DVs that originate from charm jets, a two-dimensional maximum-likelihood fit is performed to the $m_{\rm cor}({\rm DV})$ and $N_{\rm trk}({\rm DV})$ distributions, where the latter is the track multiplicity of the DV candidate. The fit procedure and the probability density functions, referred to as templates,
used to describe the charm, beauty, and light-parton  components are described in Sec.~\ref{sec:calibration:dvfits}. The requirement of the presence of a reconstructed DV in the jet, along with the application of this fit, constitutes the charm-jet tagging, or $c$-tagging, algorithm.

In simulated data, it is possible to unambiguously determine the fraction of charm jets that contain a DV candidate without the need to perform a fit. While small discrepancies are expected between data and simulation, simulated DVs should reliably reproduce the efficiency. In the simulation, the $c$-tagging efficiency is about 24\% and nearly uniform in the $20 < \ptj < 100\gev$ and $2.2 < \eta(j) < 4.2$ region.

\section{Calibration in data}

\label{sec:calibration}

The $c$-tagging efficiency is measured using a sample
of charm-enriched jets obtained from dijet events
as described in Sec.~\ref{sec:datasets}, which have been selected such that the jets remain unbiased with respect to the charm-tagging algorithm.
Exclusive charm decays are used to determine the total number of $\cquark$ jets in the sample. This is inefficient as the majority of charm hadrons do not decay to any given final state; however, the efficiency can be reliably modelled in simulation. In addition, charm-quark hadronisation into a $c$ hadron, followed by an exclusive decay, can be calculated in many cases with well known fragmentation and branching fractions.
The $\cquark$-tagging efficiency is determined as the ratio of the number of charm jets tagged to the total charm-jet yield in the sample:
\begin{equation}
\label{eq:ctag}
    \epsilon_{c{\rm -tag}} = \frac{N_{c{\rm -tag}}}{N_{\cquark}}\,,
\end{equation}
where $N_{c{\rm -tag}}$ is the number of $c$-tagged jets, {\em i.e.}\ the charm yield obtained by fitting the $[m_{{\rm cor}}({\rm DV}), N_{{\rm trk}}({\rm DV})]$ distribution for jets with an associated DV, and $N_{\cquark}$ is the total charm-jet yield.
The total $c$-jet yield is calculated separately using $\decay{\Dz}{\Km\pip}$ and $\decay{\Dp}{\Km\pip\pip}$ decays,\footnote{Note that the inclusion of charge-conjugate decay modes is implied.} collectively or generically referred to as $D$ decays hereafter, and a weighted average of the two results is used for the default $c$-tagging efficiency.
The total charm-jet yield is determined from each decay channel as
\begin{equation}
\label{eq:nctot}
    N_{\cquark}(D) = \frac{N_{\rm prompt}(D)}{\epsilon_D\, f_{\cquark\to D}\,\mathcal{B}(D)}\,,
\end{equation}
where $N_{\rm prompt}(D)$ is the observed number of promptly produced $D$ mesons obtained by fitting the \D-meson candidate mass and \chisqip distributions, $\epsilon_D$ is the efficiency with which $D$ candidates are reconstructed and selected as determined from simulation, $f_{\cquark\to D}$ is the fragmentation fraction for a charm quark to hadronise as the required $\cquark$ hadron, and $\mathcal{B}(D)$ is the corresponding branching fraction for the $D$-meson decay.
The branching and fragmentation fractions, which are derived from Refs.~\cite{PDG2020} and~\cite{Lisovyi:2015uqa}, respectively, are listed in Table~\ref{tab:bfff}.
The fragmentation fractions in Ref.~\cite{Lisovyi:2015uqa} are corrected to account for updated branching-fraction measurements~\cite{PDG2020}, and include a small correction derived from simulation for the case where ground-state $c$ hadrons produced in the decays of excited charm states, {\em e.g.}\ $D^* \to D\pi$, have $\pt(D) < 5\gev$ despite the parent state being above the 5\gev fiducial threshold.
The total correction is $\order(1\,\%)$ for both decay channels.

\begin{table}
  \begin{center}
	  \caption{\label{tab:bfff} Branching and fragmentation fractions used to obtain the total charm yields from $\decay{\Dz}{\Km\pip}$ and $\decay{\Dp}{\Km\pip\pip}$ decays.
		   The PDG~\cite{PDG2020} averages are used for both branching fractions.
		   Charm fragmentation fractions are based on the global averages reported in Ref.~\cite{Lisovyi:2015uqa}, but have been updated as detailed in the text. The fragmentation fractions are inclusive of feed down from excited charm states.}
	  \resizebox{\textwidth}{!}{
      \begin{tabular}{lcccc}
        \toprule
	Decay & Branching fraction (\%) & \multicolumn{2}{c}{Fragmentation fraction (\%)} & $\mathcal{B}(D)\times f_{c\to D}$ (\%) \\
	& & Ref.~\cite{Lisovyi:2015uqa} & Corrected &  \\
        \midrule
	$\decay{\Dz}{\Km\pip}$             & $3.950\pm0.031$ & $60.86 \pm 0.76$ & $60.12\pm0.77$ & $2.375\pm0.036$ \\
	$\decay{\Dp}{\Km\pip\pip}$         & $9.38\pm0.16$   & $24.04 \pm 0.67$ & $23.90\pm0.68$ & $2.242\pm0.074$ \\
        \bottomrule
      \end{tabular}
      }
  \end{center}
\end{table}

\subsection{Calibration datasets}
\label{sec:datasets}

\begin{figure}[t]
\centering
\includegraphics[width=0.45\columnwidth]{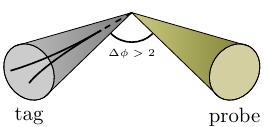}
\includegraphics[width=0.45\columnwidth]{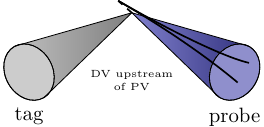}\\
\includegraphics[width=0.45\columnwidth]{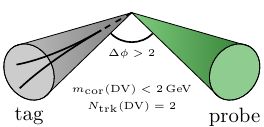}
\includegraphics[width=0.45\columnwidth]{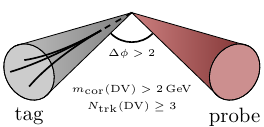}\\
\caption{Depictions of the flavour-enhanced data samples used in this analysis. The jets labelled probe are retained for further analysis. Sub-figures depict (top left) the heavy-flavour-enriched sample and (top right) the light-parton mis-tag enriched sample, as well as the further enriched (bottom left) charm and (bottom right) beauty sub-samples. Some additional requirements are applied but not included in the labeling; see text for details.}
\label{fig:dijets}
\end{figure}

The $c$-tagging efficiency is measured on a dijet control sample using a tag-and-probe method. Events are retained for further analysis if one of the two jets, henceforth called the tag jet and denoted by the symbol $j_{\rm tag}$, is associated with a DV. The jets are required to be well separated in azimuthal angle, $\Delta\phi>2$, and to have well balanced transverse momenta: $A^{jj}_{\pt}\equiv\left|\pt(j)-\pt(j_{\rm tag})\right|/\left(\pt(j)+\pt(j_{\rm tag})\right)<0.25$. Additionally, the trigger requirements introduced in Sec.~\ref{sec:Detector} are required to be fulfilled by the tag jet. These requirements enhance the fraction of heavy flavour, \ie\ $\bbbar$ and $\ccbar$, events within the sample relative to those containing light-parton jets without biasing the properties of the probe jet.
Further enriched sub-samples are also obtained by placing additional requirements on the tag-jet DV candidates. Specifically, the enriched charm-jet sub-sample requires a DV candidate with $m_{\rm cor}({\rm DV})<2\gev$ and only two tracks, while the beauty-jet sub-sample requires a DV candidate with $m_{\rm cor}({\rm DV})>2\gev$ and three or four tracks. The obtained charm-jet and beauty-jet sub-samples are approximately 30--40\% and 60--70\% pure, respectively, based on fits to the samples. These enriched sub-samples are used to perform data-driven corrections to the fit templates as described in Sec.~\ref{sec:calibration:dvfits}. An additional sample, with no DV requirements placed on the tag jet, is used to study light-parton jets. This sample is enriched in fake DVs, and hence mis-tagged light-parton jets, by retaining only DV candidates reconstructed upstream of the associated PV. The tagging requirements applied in the four samples derived from the dijet dataset are illustrated in Fig.~\ref{fig:dijets}.

\subsection{\boldmath$D$-meson decay selection and fits}
\label{sec:Dfits}

The selection and fit procedures used for $D$-meson decays closely match those used in previous studies of prompt charm production at LHCb~\cite{LHCb-PAPER-2012-041,LHCb-PAPER-2015-041,LHCb-PAPER-2016-042}.
The $D$-meson decays are reconstructed by combining good-quality charged tracks with requirements placed on their momentum, $\pt$, and $\chisqip$.  Kaon candidates are also required to either pass a kaon particle-identification requirement or have high momentum. Requirements are also placed on the invariant mass and $\pt$ of the combination as well as the vertex quality, the significance of separation from the PV, and the angle between the flight direction from the PV and the momentum vector. For $\Dz$ candidates, a requirement is also placed on the distance of closest approach of the two charged particle tracks. For $\Dp$ candidates, an additional requirement is placed on the minimum decay time. All candidates are required to have momentum vectors that fall within $\Delta R<0.5$ of the jet axis.

To distinguish promptly produced charm hadrons from those produced in $\bquark$-hadron decays and from combinatorial background, a two-dimensional unbinned maximum-likelihood fit is performed to the invariant mass and $\log\chisqip$ distributions of the $D$-meson candidates. The mass distributions of the prompt and from-$\bquark$ components are each described by the sum of a Gaussian with a Crystal Ball function~\cite{Skwarnicki:1986xj}, while the background is described by a linear function. The $\log\chisqip$ distributions of the prompt and from-$\bquark$ components are described by asymmetric Gaussian functions with exponential tails, while the background is described by a kernel density estimation derived from data in the mass-sideband regions. Various shape parameters of the fit components are fixed to values determined from simulation. To better describe the data, fits are performed simultaneously to five intervals of $\pt(D)/\pt(j)$, with some shape parameters allowed to vary independently in each interval.

\begin{figure}[t]
\centering
\includegraphics[width=0.45\columnwidth]{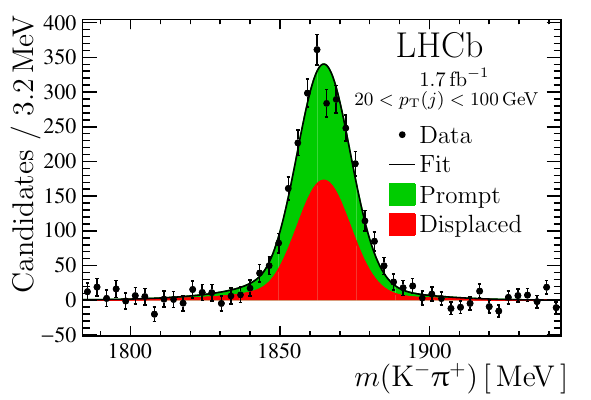}
\includegraphics[width=0.45\columnwidth]{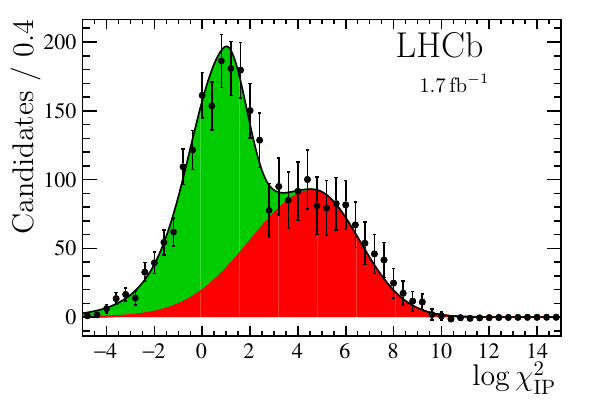}\\
\includegraphics[width=0.45\columnwidth]{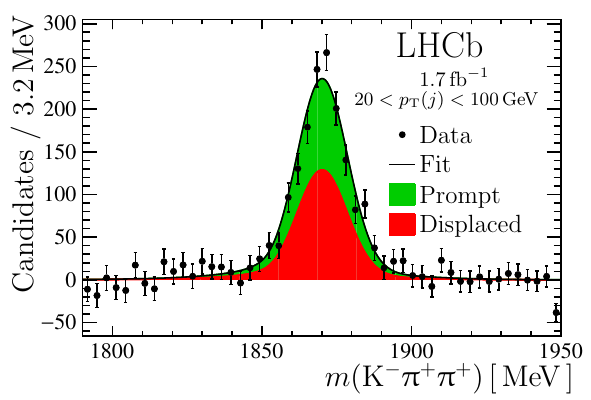}
\includegraphics[width=0.45\columnwidth]{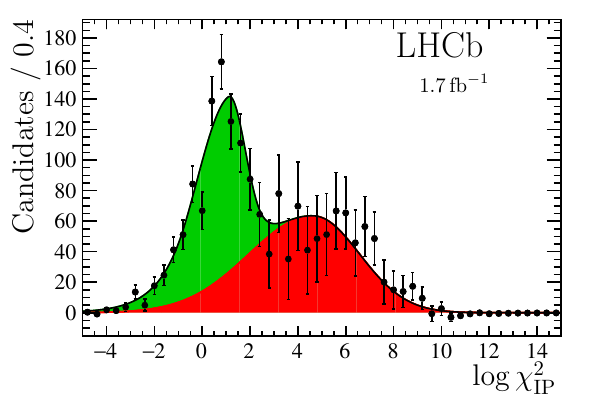}\\
\caption{Background-subtracted (left) invariant mass and (right) $\log\chisqip$ projections and fit results for all (top) $\decay{\Dz}{\Km\pip}$ and (bottom) $\decay{\Dp}{\Km\pip\pip}$ candidates associated with jets reconstructed in the efficiency-reporting region. The background uncertainties, which are included in the error bars, predominantly affect the displaced components.}
\label{fig:dfit}
\end{figure}

Figure~\ref{fig:dfit} shows the combinatorial-background-subtracted invariant mass and $\log\chisqip$ distributions for all $\Dz$ and $\Dp$ candidates associated with jets reconstructed in the efficiency-reporting region, along with projections of fits performed on these samples. Such fits are performed in each interval of jet $\pt$. The prompt signal yields extracted from these fits are scaled by an efficiency-correction factor, which is determined from simulated events as a function of the charm-hadron kinematics, and weighted according to the kinematic distribution of candidates in the signal region of invariant mass and $\log\chisqip$.

\subsection{Displaced vertex fits}
\label{sec:calibration:dvfits}

\begin{figure}[t]
\centering
\includegraphics[width=0.45\columnwidth]{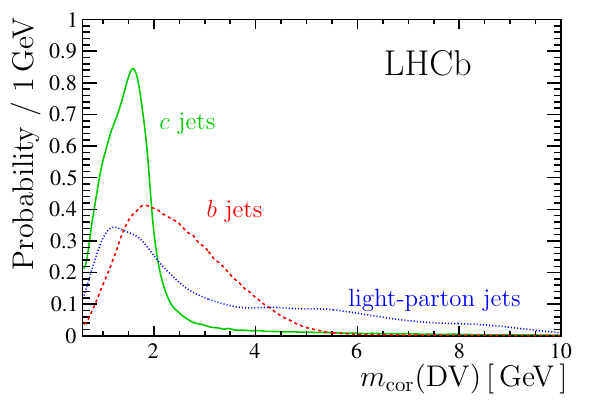}
\includegraphics[width=0.45\columnwidth]{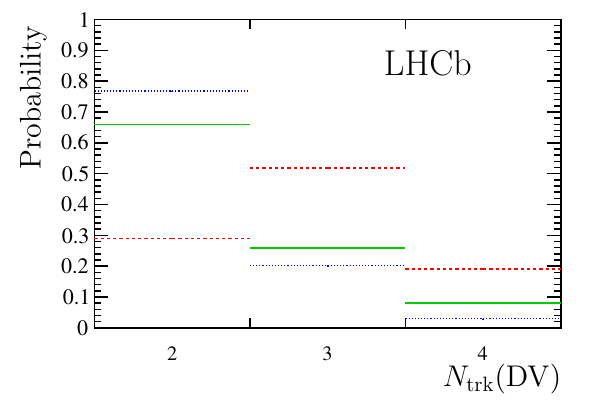}\\
\caption{Probability density functions for (left) $m_{\rm cor}({\rm DV})$ and (right) $N_{\rm trk}({\rm DV})$ used in the fits for (solid green) charm, (dashed red) beauty, and (dotted blue) light-parton jets. }
\label{fig:svtemplates}
\end{figure}

\begin{figure}[t]
\centering
\includegraphics[width=0.45\columnwidth]{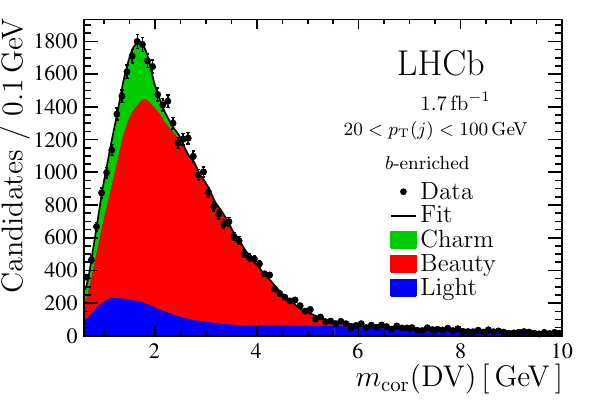}
\includegraphics[width=0.45\columnwidth]{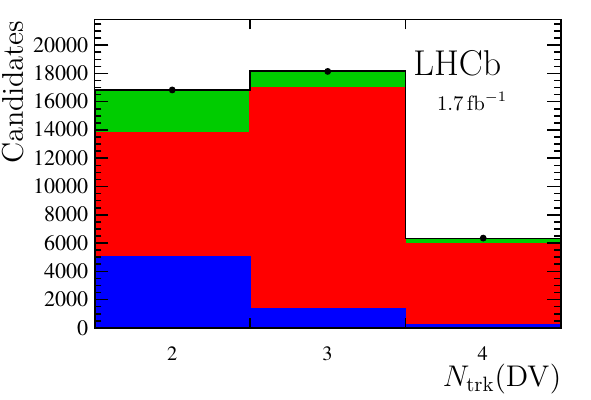}\\
\includegraphics[width=0.45\columnwidth]{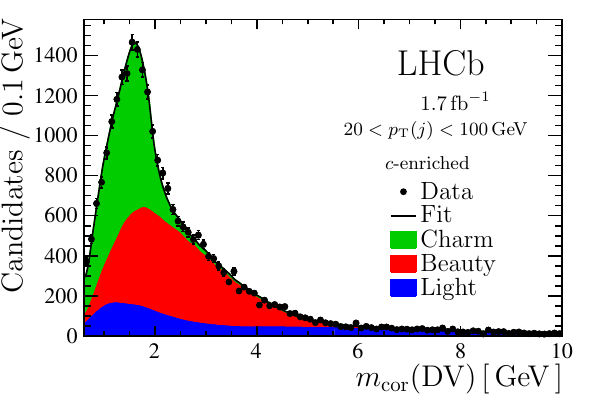}
\includegraphics[width=0.45\columnwidth]{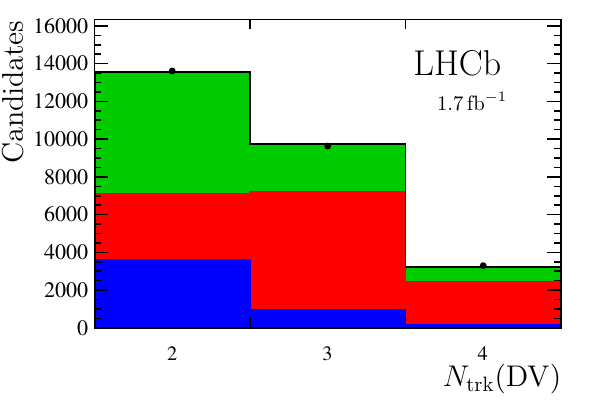}\\
\includegraphics[width=0.45\columnwidth]{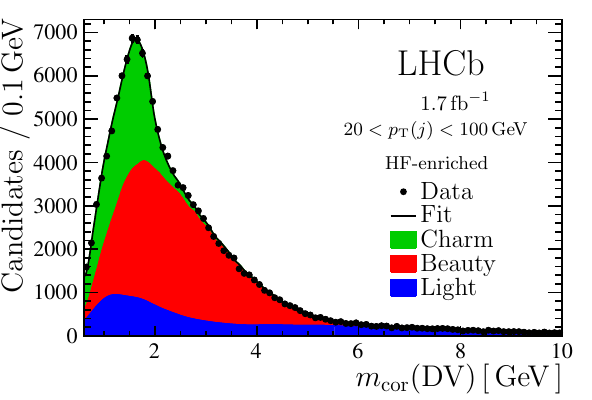}
\includegraphics[width=0.45\columnwidth]{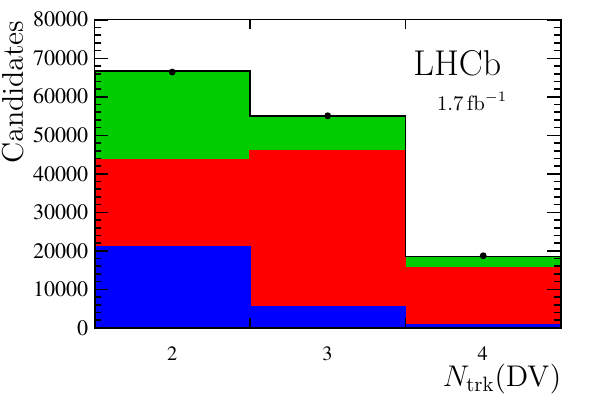}
\caption{DV (left) corrected mass and (right) track multiplicity projections of fits to the flavour-enriched jet samples: (top-to-bottom)
beauty-enriched sub-sample, charm-enriched sub-sample, and
heavy-flavour-enriched sample fit with data-driven corrections.}
\label{fig:svfit}
\end{figure}

Candidate DVs are selected as described in Sec.~\ref{sec:identification}. A  two-dimensional fit is performed to the corrected mass and track multiplicity distributions to extract the $\cquark$-jet component. The template describing the light-parton-jet background is taken from jets with the DV displacement requirement reversed, such that reconstructed DV candidates are displaced backwards with respect to the PV. Templates describing the distributions of the $\cquark$ and $\bquark$ components are taken from simulation but corrected to match data using fits performed to subsets of the data that have been further enriched in charm and beauty as described in Sec.~\ref{sec:datasets}. After each fit, the template of the enriched component is modified to minimise the residuals. Fits are first performed to the enriched beauty sample and then to the enriched charm sample, and this process is repeated iteratively until the fit results change by less than 1\%. In practice, a single iteration is found to be sufficient. Templates for the $m_{\rm cor}({\rm DV})$ and $N_{\rm trk}({\rm DV})$ distributions are shown in Fig~\ref{fig:svtemplates}. Examples of $m_{\rm cor}({\rm DV})$ and $N_{\rm trk}({\rm DV})$ distributions together with fit projections to the charm- and beauty-enriched sub-samples and to the full heavy-flavour-enriched sample with corrections applied are shown in  Fig.~\ref{fig:svfit}.
As this study defines charm jets as those containing a $\cquark$ hadron with $\pt>5\gev$, a correction must be applied to account for cases where a $\pt<5\gev$ $c$ hadron produces a DV candidate. However, this correction is found to be $\mathcal{O}(1\%)$ in all of the $\pt(j)$ ranges considered.
Furthermore, this correction largely cancels when using the $c$-tagging efficiencies measured here to correct $c$-jet yields in analyses that employ the same charm-jet definition.

\subsection{Unfolding}
\label{sec:unfolding}

\begin{figure}[t]
\centering
\includegraphics[width=0.45\columnwidth]{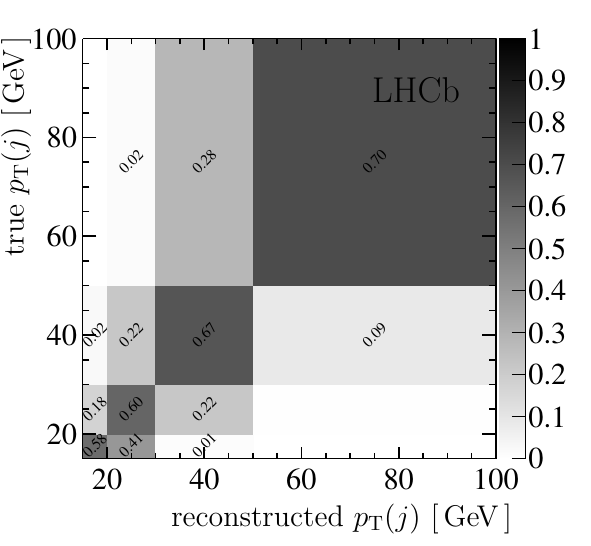}
\includegraphics[width=0.45\columnwidth]{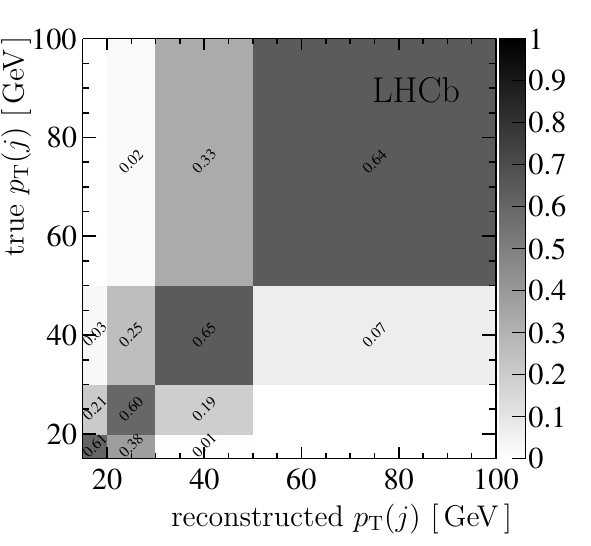}\\
\includegraphics[width=0.45\columnwidth]{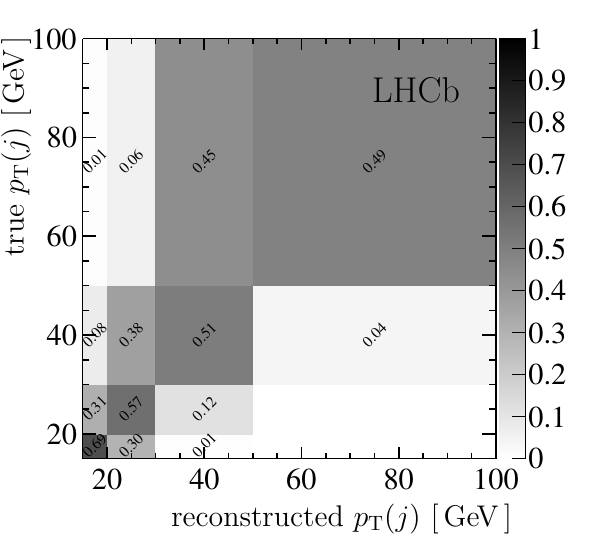}
\caption{Detector-response matrices for (left) $\Dz$-, (right) $\Dp$- and (bottom) DV-tagged charm jets. The shading represents the interval-to-interval migration probabilities ranging from (white) 0 to (black) 1 such that each row sums to unity when the underflow and overflow bins are included.
Jets with true (reconstructed) \ptj in the 20--100\gev region but whose reconstructed (true) \ptj is either below 15\gev or above 100\gev are included in the unfolding but not shown graphically.}
\label{fig:unfold}
\end{figure}

As \ptj resolution effects may differ between jets containing a reconstructed $D$-meson decay or a DV candidate, $\pt(j)$ interval migration must be considered separately for the numerator and denominator in the $c$-tagging efficiency measurement. In both cases, unfolding is performed using an iterative Bayesian procedure~\cite{D'Agostini:1994zf} as implemented in {\sc RooUnfold}\cite{Adye:2011gm} with two iterations. The unfolding matrices for DV-tagged charm jets as well as charm jets containing reconstructed $\Dz$ and $\Dp$ decays are shown in Fig.~\ref{fig:unfold}. These are determined from simulated data that have been weighted to better describe the $\pt(j)$, $\pt({\rm DV})$, and $\pt(D)$ distributions observed in data.
In addition, the detector response is studied in data using the \pt-balance distribution $\ptj/\pt(Z)$ for $\Z+$jet candidates that are nearly back-to-back in the transverse plane, using the same technique as in Refs.~\cite{LHCb-PAPER-2013-058,LHCb-PAPER-2016-064}.
Small adjustments are applied to the \ptj scale and resolution in simulation to obtain the best agreement with data.

\subsection{Systematic uncertainties}
\label{sec:syst}

Two categories of systematic uncertainty affect the $c$-tagging efficiency: those that affect both the efficiency measurement performed here and the $c$-tagged data samples used in subsequent measurements, \eg\ uncertainties in the DV fitting procedure; and those that affect only the $c$-tagging efficiency, \eg\ the $D$-meson fitting procedure. The former category of uncertainties partially cancel in any efficiency-corrected results, \eg\ the measurement of $\sigma(Zc)/\sigma(Zj)$~\cite{LHCb-PAPER-2021-029}, and  therefore must be calculated separately for each study that uses this tagging method. Sources of uncertainty that fall into the latter category, which do not cancel in analyses, are considered below.

The systematic uncertainty due to the $D$-decay $[m(\Km\pip),\log\chisqip]$ and $[m(\Km\pip\pim),\log\chisqip]$ fits  accounts for imperfect knowledge of the probability density functions used to model the three fit components. The uncertainty is assigned as the largest deviation from the default results based on several variations to the fit model.
An uncertainty is assigned due to the efficiency-weighting procedure applied to the $D$-decay yields. Specifically, the integrated efficiency factor is replaced by a per-event efficiency correction and the differences from the default results are assigned as systematic uncertainties.
Additional systematic uncertainties are assigned due to the limited size of simulation samples used to determine $D$-decay efficiencies, the procedure used to determine the efficiency of particle-identification requirements, and potential data--simulation discrepancies in modeling the detector response.
Uncertainties on the fragmentation  and branching fractions  are propagated through to the $c$-tagging efficiency results accounting for correlations.
The uncertainties for each source are listed in Table~\ref{tab:systs}.

\begin{table}
  \begin{center}
	  \caption{\label{tab:systs}
	    Relative systematic uncertainties (\%) on the tagging efficiencies determined using the $\Dz$ and $\Dp$ decays as well as their weighted combination. Ranges of uncertainties are given when the value depends on the $\ptj$ interval. The total systematic uncertainty is evaluated as the sum in quadrature of the uncertainties from all sources.}
      \begin{tabular}{lccc}
        \toprule
  Source  & \multicolumn{3}{c}{Uncertainty (\%)} \\
  & $\Dz$    & $\Dp$    & Combination \\
\midrule
$D$ fit models &   4 &   5--18 &   3--6\\
$D$ efficiency method &   1--2 &   3--8 &   1--2\\
Simulation sample size &   1 &   2--4 &   1\\
Particle identification &   1--2 &   4--7 &   1--2\\
Modeling detector response &   2 &   2 &   2\\
Fragmentation \& branching fractions &   2 &   3 &   1\\
2015-16 {\em vs} 2017-18 &   2 &   2 &   2\\
\midrule
Total &   5--6 &   9--21 &   5--7\\
        \bottomrule
      \end{tabular}
  \end{center}
\end{table}

As the error parameterisation used during track reconstruction was changed between data collected in 2015-16 and 2017-18, the tagging efficiency differs between these datasets. This leads to a need for a correction to the tagging efficiencies calculated from this study, which uses only data recorded in 2016. The impact of this change on the $c$-tagging efficiency is obtained from simulation.
After accounting for the relative proportions of 2015-16 and 2017-18 data in the full \lhcb Run 2 dataset, a 2\% correction to the $c$-tagging efficiency is obtained, which is also assigned as a systematic uncertainty.

As discussed above, systematic uncertainties that affect both the $c$-tagging efficiency measurement and the $c$-tagged data samples used in subsequent measurements partially cancel in any efficiency-corrected results. Therefore, these uncertainties must be estimated separately for each measurement and are not reported here.
Additional measurement-dependent sources of uncertainty include the unfolding procedure, jet reconstruction efficiency, jet energy scale and resolution, and the DV-fit templates. In Ref.~\cite{LHCb-PAPER-2021-029}, these sources contribute an additional 4--5\% relative uncertainty to the results.

\section{Results}

The $c$-tagging efficiency is measured in intervals of $\pt(j)$ using Eq.~\eqref{eq:ctag}, {\em i.e.}\ as the ratio of the number of DV-tagged charm jets to the total charm-jet yield in the control sample.
The DV-tagged charm yields in intervals of reconstructed $\pt(j)$ are obtained from the fits described in Sec.~\ref{sec:calibration:dvfits}.
The total charm-jet yields, also in intervals of reconstructed $\pt(j)$, are obtained using Eq.~\eqref{eq:nctot}, which takes as input the $D$-meson fit results and efficiency corrections of Sec.~\ref{sec:Dfits}, and the fragmentation and branching fractions from Table~\ref{tab:bfff}.
Interval migration due to $\pt(j)$ resolution is accounted for separately for the DV-tagged and total charm-jet yields using the unfolding approach of Sec.~\ref{sec:unfolding}.

\begin{figure}[t]
\centering
\includegraphics[width=0.9\columnwidth]{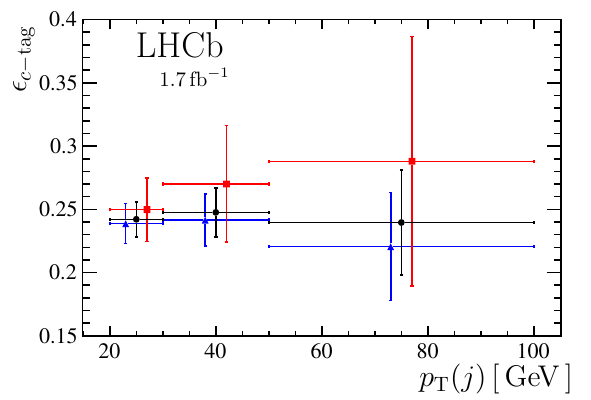}
\caption{Charm-tagging efficiency in intervals of \pt determined from (blue triangles) $\decay{\Dz}{\Km\pip}$ and (red squares) $\decay{\Dp}{\Km\pip\pip}$ decays, as well as (black circles) the weighted average. The points are offset in each \pt interval to aid visibility.
}
\label{fig:tageffcomb}
\end{figure}

The measured Run~2 tagging efficiency is given in Table~\ref{tab:results} in intervals of $\ptj$ as well as integrated in $\ptj$.
Comparing to simulation, the scale factors required to correct the $c$-tagging efficiency are determined to be $1.03 \pm 0.06$, $1.01 \pm 0.08$, and $1.09 \pm 0.17$ in the 20--30, 30--50, and 50--100\gev \ptj intervals, respectively, which include both the statistical and systematic uncertainties. As described in Sec.~\ref{sec:syst}, systematic uncertainties that affect both the efficiency measurement and the $c$-tagged data samples used in subsequent measurements are not included because these will largely cancel.
Figure~\ref{fig:tageffcomb} displays the $c$-tagging efficiencies and compares the results obtained from the two $D$-meson decays separately.
For the measurement presented in Ref.~\cite{LHCb-PAPER-2021-029},  which involved integrating over \ptj, the relative $c$-tagging efficiency uncertainty is 6\% including both the statistical and systematic contributions.

\begin{table}
  \begin{center}
	  \caption{\label{tab:results}
	    Charm-tagging efficiencies (\%) determined in intervals of $\ptj$. First and second uncertainties are statistical and systematic, respectively. Note that systematic uncertainties that affect both the efficiency measurement and the $c$-tagged data samples used in subsequent measurements are not included because these will largely cancel.}
      \begin{tabular}{cccc}
        \toprule
	\multicolumn{4}{c}{\ptj interval [\gev]} \\
$(20,30)$ & $(30,50)$ & $(50,100)$ & $(20,100)$\\
        \midrule
$23.9 \pm 0.7 \pm 1.2$
& $24.4 \pm 1.4 \pm 1.3$
& $23.6 \pm 3.7 \pm 1.7$
& $24.0 \pm 0.6 \pm 1.4$\\
        \bottomrule
      \end{tabular}
  \end{center}
\end{table}

\section{Summary}
\label{sec:summary}

In summary, the identification of charm jets is achieved at LHCb in Run~2 using a method based on the properties of displaced vertices reconstructed within the jets. The performance of this method is determined using an unbiased dijet calibration dataset recorded by the LHCb detector during the same data-taking period. The charm-tagging efficiency in data, which is found to be consistent with simulation, is reported as a function of the transverse momentum of the jet,  and found to be about 24\% for $20 < \ptj < 100\gev$ and $2.2 < \eta(j) < 4.2$.

\section*{Acknowledgements}
\noindent We express our gratitude to our colleagues in the CERN
accelerator departments for the excellent performance of the LHC. We
thank the technical and administrative staff at the LHCb
institutes.
We acknowledge support from CERN and from the national agencies:
CAPES, CNPq, FAPERJ and FINEP (Brazil);
MOST and NSFC (China);
CNRS/IN2P3 (France);
BMBF, DFG and MPG (Germany);
INFN (Italy);
NWO (Netherlands);
MNiSW and NCN (Poland);
MEN/IFA (Romania);
MSHE (Russia);
MICINN (Spain);
SNSF and SER (Switzerland);
NASU (Ukraine);
STFC (United Kingdom);
DOE NP and NSF (USA).
We acknowledge the computing resources that are provided by CERN, IN2P3
(France), KIT and DESY (Germany), INFN (Italy), SURF (Netherlands),
PIC (Spain), GridPP (United Kingdom), RRCKI and Yandex
LLC (Russia), CSCS (Switzerland), IFIN-HH (Romania), CBPF (Brazil),
PL-GRID (Poland) and NERSC (USA).
We are indebted to the communities behind the multiple open-source
software packages on which we depend.
Individual groups or members have received support from
ARC and ARDC (Australia);
AvH Foundation (Germany);
EPLANET, Marie Sk\l{}odowska-Curie Actions and ERC (European Union);
A*MIDEX, ANR, IPhU and Labex P2IO, and R\'{e}gion Auvergne-Rh\^{o}ne-Alpes (France);
Key Research Program of Frontier Sciences of CAS, CAS PIFI, CAS CCEPP,
Fundamental Research Funds for the Central Universities,
and Sci. \& Tech. Program of Guangzhou (China);
RFBR, RSF and Yandex LLC (Russia);
GVA, XuntaGal and GENCAT (Spain);
the Leverhulme Trust, the Royal Society
 and UKRI (United Kingdom).

\addcontentsline{toc}{section}{References}
\ifx\mcitethebibliography\mciteundefinedmacro
\PackageError{LHCb.bst}{mciteplus.sty has not been loaded}
{This bibstyle requires the use of the mciteplus package.}\fi
\providecommand{\href}[2]{#2}

\newpage

\centerline
{\large\bf LHCb collaboration}
\begin
{flushleft}
\small
R.~Aaij$^{32}$,
A.S.W.~Abdelmotteleb$^{56}$,
C.~Abell{\'a}n~Beteta$^{50}$,
F.J.~Abudinen~Gallego$^{56}$,
T.~Ackernley$^{60}$,
B.~Adeva$^{46}$,
M.~Adinolfi$^{54}$,
H.~Afsharnia$^{9}$,
C.~Agapopoulou$^{13}$,
C.A.~Aidala$^{87}$
% ...

{\footnotesize \it

% ...
$^{9}$Universit{\'e} Clermont Auvergne, CNRS/IN2P3, LPC, Clermont-Ferrand, France\\
% ...
$^{13}$LPNHE, Sorbonne Universit{\'e}, Paris Diderot Sorbonne Paris Cit{\'e}, CNRS/IN2P3, Paris, France\\
% ...
$^{32}$Nikhef National Institute for Subatomic Physics, Amsterdam, Netherlands\\
% ...
$^{46}$Instituto Galego de F{\'\i}sica de Altas Enerx{\'\i}as (IGFAE), Universidade de Santiago de Compostela, Santiago de Compostela, Spain\\
% ...
$^{50}$Physik-Institut, Universit{\"a}t Z{\"u}rich, Z{\"u}rich, Switzerland\\
% ...
$^{54}$H.H. Wills Physics Laboratory, University of Bristol, Bristol, United Kingdom\\
% ...
$^{56}$Department of Physics, University of Warwick, Coventry, United Kingdom\\
% ...
$^{60}$Oliver Lodge Laboratory, University of Liverpool, Liverpool, United Kingdom\\
% ...
$^{68}$Syracuse University, Syracuse, NY, United States\\
% ...
$^{87}$University of Michigan, Ann Arbor, United States, associated to $^{68}$\\
\bigskip
% ...
}
\end{flushleft}

\end{document}